\documentclass[12pt]{article}
\usepackage[leqno]{amsmath}
\usepackage{latexsym,epic,eepic,pstcol,comment}
\usepackage{graphics}

\smallskip
\newcommand\QED{\hfill\raisebox{1pt}{\framebox[2.5mm][r]{
$\phantom x$}}\bigskip}

\newcommand{\Number}[1]{\mathbf{#1}}

\newcommand{\Transformation}[1]{\mathsf{#1}}

\newcommand{\C}{\Number{C}}

\newcommand{\R}{\Number{R}}

\newcommand{\eps}{\varepsilon}
\newcommand{\Action}{\Transformation{R}}
\newcommand{\Wick}{\Transformation{W}}

\newcommand{\Angle}{\psi}
\newcommand{\cyl}{\Phi}
\newcommand{\curv}{K}
\newcommand{\grav}{\kappa}
\newcommand{\rad}{\rho}

\DeclareMathOperator{\sech}{sech}

\newcommand{\sig}{{1+1}} 
\newcommand{\Sig}{{2+1}} 

\DeclareMathOperator{\re}{Re}

\newcommand{\lapl}{\lower 1pt\hbox{$\Box$}}

\newcommand{\profile}{\phi}

\newcommand{\dual}[1]{\widehat{#1}}

\newcommand{\Neg}{\phantom{-}}

\newcommand{\be}{\begin{enumerate}}
\newcommand{\ee}{\end{enumerate}}
\newcommand{\bi}{\begin{itemize}}
\newcommand{\ei}{\end{itemize}}
\newcommand{\bd}{\begin{description}}
\newcommand{\ed}{\end{description}}
\newcommand{\bt}{\begin{tabular}}
\newcommand{\et}{\end{tabular}}

\title{Spacetime Slices and Surfaces of Revolution}
\author{John T. Giblin and Andrew D. Hwang}
\date{}

\begin{document}

\maketitle

\begin{abstract}
  Under certain conditions, a $(\sig)$-dimensional slice~$\dual{g}$ of
  a spherically symmetric black hole spacetime can be equivariantly
  embedded in $(\Sig)$-dimensional Minkowski space. The embedding
  depends on a real parameter that corresponds physically to the
  surface gravity~$\grav$ of the black hole horizon.
  
  Under conditions that turn out to be closely related, a real surface
  that possesses rotational symmetry can be equivariantly embedded in
  3-dimensional Euclidean space. The embedding does \emph{not}
  obviously depend on a parameter. However, the Gaussian curvature is
  given by a simple formula: If the metric is written
  $g = \profile(r)^{-1}\,dr^2 + \profile(r)\,d\theta^2$,
  then $\curv_g=-\frac{1}{2}\profile''(r)$.
  
  This note shows that metrics $g$~and $\dual{g}$ occur in dual pairs,
  and that the embeddings described above are orthogonal
  facets of a single phenomenon. In particular, the metrics and their
  respective embeddings differ by a Wick rotation that preserves the
  ambient symmetry. 
  
  Consequently, the embedding of~$g$ depends on a real parameter.  The
  ambient space is not smooth, and~$\grav$ is inversely proportional
  to the cone angle at the axis of rotation.  Further, the Gaussian
  curvature of~$\dual{g}$ is given by a simple formula that seems not
  to be widely known.
\end{abstract}

\section{Introduction}
\label{section:intro}

The most concrete way to study a surface is
(when possible) to embed it isometrically in a flat 3-dimensional space.
Isometric embedding ``realizes'' the abstract surface, and is useful
for developing geometric intuition.  Naturally, the signature of the
surface metric is related to the signature of the ambient metric. For
example, a ``spacetime slice'' of signature~$(\sig)$ cannot be
isometrically embedded in~$\R^3$; instead, one might seek an embedding
into the Minkowski space~$\R^\Sig$.

When an abstract surface has symmetry, it is natural to seek an
embedding in which the intrinsic symmetry is realized by a symmetry
of the ambient space.  When this occurs, the embedding is said to be
\emph{equivariant} (with respect to the actions of the abstract
symmetry groups).

This note originates with two families of equivariant isometric
embeddings. The first realizes slices of certain spherically-symmetric
static spacetimes as surfaces in Minkowski space~$\R^\Sig$,
see~\cite{GMG, ma}. (These papers were in turn inspired by earlier
work~\cite{Davidson, Deser, Fronsdal, Kasner, Pad, Pad1} involving
higher-dimensional embeddings into flat space.) The second is a
``symplectic'' description of (classical) surfaces of revolution,
see~\cite{H}.  Each family of metrics has continuous symmetry: time
translation and rotation, respectively.

These two families of metrics are equivariantly related by ``Wick
rotation'', in a sense made precise below.  The relationship is
roughly analogous to the duality between conjugate minimal surfaces:
For each suitable spacetime slice, there is a ``dual'' surface of
revolution, and a single equivariant embedding into~$\C^3$ that
interpolates the embeddings of the respective surfaces into flat
(real) 3-dimensional spaces. The resulting 1-parameter family of
metrics on the abstract surface interpolates metrics of different
signatures, so unlike the situation for minimal surfaces, the
interpolating metrics are not mutually isometric (even locally).

Geometric properties of one family have consequences for the other
family:

\bi
\item The Gaussian curvature of a spacetime slice is given by an
  extremely simple formula.

\item The surface gravity of a black hole horizon has a geometric
  interpretation as angular defect of a cone singularity. In
  particular, the metrics studied in~\cite{H} are instances of a
  1-parameter family of (possibly singular) metrics that are
  ``smoothly embedded'' in a (flat) singular ambient space.
\ei

\subsection*{Organization}

In Section~\ref{section:intrinsic} we introduce the abstract surfaces
under consideration; fixed points are removed for simplicity, and the
intrinsic ``Wick rotation'' is described as a discrete symmetry. In
Section~\ref{section:ambient}, we equip the ambient space~$\C^3$ with
coordinates, an action of the additive complex group, and an
appropriate (flat) indefinite metric. Section~\ref{section:embeddings}
describes the interpolating embedding, a map from $\C\times\R$
to~$\C^3$. The equivariant embeddings of spacetime slices and surfaces
of revolution are the restrictions of this embedding to
$i\R\times\R$~and $\R\times\R$. In Section~\ref{section:boundary}, we
address ``boundary'' questions of smoothness and embeddability.

\section{Intrinsic Metrics}
\label{section:intrinsic}
\setcounter{equation}{0}
\setcounter{figure}{0}


We take the mathematical point of view that a ``metric'' comprises a
metric tensor \emph{and} a parameter domain (or manifold). Thus, a
metric can be 2-dimensional, compact, connected, oriented, etc.

Consider an oriented 2-dimensional metric that admits a free,
isometric action of the additive group~$\R$. The interpretation of
the symmetry depends on the signature: In Lorentz signature, the
action should be viewed as time translation of a $(\sig)$-dimensional
slice of a static spacetime. In Riemannian signature, the action is
something like a rotation about an axis, except that fixed points have
been removed and the resulting metric lifted to its universal cover.

The existence of useful coordinate systems is independent of the
metric signature, so for the moment we work with a Riemannian
metric~$g$.

\subsection*{Isothermal Parameters}

Away from fixed points of the action, existence of isothermal
parameters is elementary: Construct nets of curves by taking orbits of
the action and the associated $g$-orthogonal family. Clearly these
curves are coordinate curves in which the metric is diagonal and the
group action is a coordinate translation. A change of variable in the
direction transverse to the action ensures that the diagonal metric
components are equal; see~\cite{H} for details.

More precisely, our metric may be regarded as living on the planar
domain $D:=\R\times(a,b)$ with coordinates~$(t,s)$, and with~$\R$
acting by translation in~$t$, Figure~\ref{fig:domain}.  After
changing~$s$ if necessary, there exists a smooth function
$\varphi:(a,b)\to\R$ such that
\begin{equation}
  \label{eqn:metric_s}
  g = \varphi(s)(dt^2 + ds^2).
\end{equation}
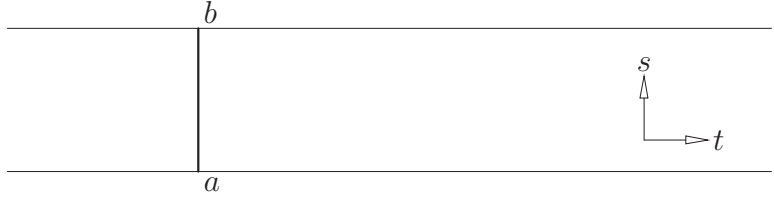
\begin{figure}[hbt]
\begin{center}
\setlength{\unitlength}{1in}
\begin{picture}(4,1)(-0,-0)
\path(3.33333,0.333333)(3.55251,0.333333)
\path(3.55251,0.354089)(3.55251,0.312578)(3.66667,0.333333)(3.55251,0.354089)
\path(3.33333,0.333333)(3.33333,0.552511)
\path(3.31258,0.552511)(3.35409,0.552511)(3.33333,0.666667)(3.31258,0.552511)
\path(0,0.166667)(4,0.166667)
\path(0,0.916667)(4,0.916667)
\thicklines
\path(1,0.166667)(1,0.916667)
\put(1.02767,0.138993){\makebox(0,0)[tl]{$a$}}
\put(1.02767,0.944341){\makebox(0,0)[bl]{$b$}}
\put(3.69434,0.333333){\makebox(0,0)[l]{$t$}}
\put(3.33333,0.694341){\makebox(0,0)[b]{$s$}}
\end{picture}
\caption{The domain of an intrinsic metric}
\label{fig:domain}
\end{center}
\end{figure}

\subsection*{Action-Angle Coordinates}

An additional change of transverse coordinate expresses the metric in
\emph{action-angle} (or \emph{symplectic}) form
\begin{equation}
\label{eqn:metric_r}
  g = \profile(r)\,dt^2 + \frac{1}{\profile(r)}\,dr^2.
\end{equation}
Indeed, set
\begin{equation}
  \label{eqn:coord}
  r=\int^s \varphi(\zeta)\,d\zeta,\qquad
  \profile(r)=\varphi(s).
\end{equation}
The integral equation defines~$r$ as a function of~$s$ (up to an
additive constant) on each interval where~$\varphi$ is non-vanishing,
and the resulting coordinate is easily shown to
satisfy~\eqref{eqn:metric_r}. The additive constant amounts to
translation of the interval~$(a,b)$, and is geometrically harmless.

Given a non-vanishing function $\profile:(a,b)\to\R$, the equations
\begin{equation}
  \label{eqn:coord_inverse}
  s=\int^r \frac{d\zeta}{\profile(\zeta)},\qquad
  \varphi(s)=\profile(r),
\end{equation}
define a translation-invariant, isothermal metric on~$D$ that
satisfies~\eqref{eqn:metric_s}. These constructions are inverse to
each other up to isometry of~$g$ and domain translation of~$\profile$.
In other words, an isometry class of~$g$ corresponds to a
function~$\profile$ that is unique up to translation in the domain.
For reasons originating in symplectic geometry, this correspondence is
called the \emph{momentum construction}, and the function~$\profile$
is called the \emph{momentum profile} of~$g$.

\subsection*{Elementary Metric Geometry}

In action-angle coordinates, the arc length element along a generator,
and the area form, are
\begin{equation}
  \label{eqn:length}
  d\sigma=\frac{dr}{\sqrt{\profile(r)}},\qquad
  dA=dr\,dt.
\end{equation}
The second formula highlights the geometric significance of the
coordinate~$r$: In a region of surface defined by the inequalities
$c_1\leq t\leq c_2$, the change in~$r$ is proportional to the enclosed
surface area. On a surface of revolution where the $t$-interval
$[0,2\pi]$ corresponds to one full turn, $2\pi r$~measures zonal
area.

The Killing field~$\frac{d}{dt}$ that generates the group action has
squared length~$\profile(r)$. This formula is meaningful even for
metrics of Lorentz signature: The symmetry is time translation! For a
Riemannian metric embedded as a surface of revolution in~$\R^3$, one
full turn of the surface is a $t$-interval of length~$2\pi$,
and~$\sqrt{\profile(r)}$ is the Euclidean radius of the surface
at~$r$.

\subsection*{Gaussian Curvature}

The interplay between action-angle and isothermal (i.e., holomorphic)
coordinates on a surface of revolution leads to a remarkable formula
for the Gaussian curvature, see~\cite{H}:
\[
\curv=-\frac{1}{2}\profile''(r).
\]
On the Lorentzian side, direct calculation shows that the Gaussian
curvature of a spacetime slice is
\[
\curv=\frac{1}{2}\profile''(r).
\]
In both signatures
\begin{equation}
  \label{eqn:curvature}
  \curv=-\frac{1}{2}g_{tt}''(r).
\end{equation}

From the standpoint of constructing metrics of specified (e.g.,
constant) curvature, action-angle coordinates have the substantial
advantage that the Gaussian curvature is a \emph{linear} function of
the profile. Further, as we shall see, metrics presented in this form
are explicitly embeddable in a flat ambient space.

\subsection*{Complexification}

Regard~$t$ as a \emph{complex} coordinate. The complex-valued tensor
\[
  g_\C = \profile(r)\,dt^2 + \frac{1}{\profile(r)}\,dr^2
\]
on $\C\times(a,b)$ is real-valued on $\R\times(a,b)$, where it is
positive-definite, and on~$i\R\times(a,b)$, where it has Lorentz
signature.  \emph{Intrinsic Wick rotation} is the transformation
$t\mapsto it$, under which~$g$ corresponds to
\begin{equation}
\label{eqn:metric_r_dual}
  \dual{g} =  - \profile(r)\,dt^2 + \frac{1}{\profile(r)}\,dr^2.
\end{equation}
We assume from now on that the additive group~$\C$ acts by translation
in~$t$.  Since~$t$ is determined by the group action, while~$r$ is
characterized by~\eqref{eqn:metric_r}, the duality
$g\leftrightarrow\dual{g}$ (which is defined using coordinates)
actually depends only on~$g_\C$. The metric~$\dual{g}$ is the
\emph{spacetime slice} associated to the \emph{surface of
  revolution}~$g$.

\subsection*{An Example}

\begin{figure}[htb]
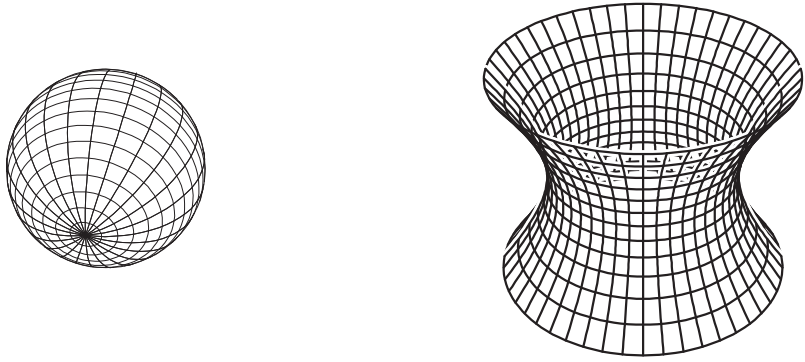

\begin{center}
\input{round1.eepic}
\qquad\qquad
\input{desitter.eepic}
\caption{Riemannian and Lorentzian embeddings of $\profile(r)=1-r^2$.}
\label{fig:embedding0}
\end{center}
\end{figure}

Consider the momentum profile $\profile(r)=1-r^2$ on~$(-1,1)$.
By~\eqref{eqn:coord_inverse},
\[
s=\int_0^r\frac{d\zeta}{1-\zeta^2}
 =\frac{1}{2}\log\Big|\frac{1+r}{1-r}\Big|,
\]
or $r=\tanh s$ and $\varphi(s)=\sech^2 s$. The arc length element and
function are
\[
d\sigma=\frac{dr}{\sqrt{1-r^2}},\qquad
\sigma=\arcsin r, \text{ or } r=\sin\sigma.
\]
If this metric is embedded isometrically in~$\R^3$, the radius at~$r$
is $\sqrt{\profile(r)}=\sqrt{1-r^2}=\cos\sigma$. As should be clear,
we are looking at the round metric of unit radius on a sphere, a fact
that is confirmed (at least circumstantially) by the fact that the
Gaussian curvature is $-\frac{1}{2}\profile''(r)=1$.  The
dual spacetime slice is
\[
\dual{g}=\frac{dr^2}{1-r^2}-(1-r^2)\,dt^2,
\]
a portion of the well-known deSitter metric, which embeds in~$\R^\Sig$
as (part of) a hyperboloid of one sheet. Figure~\ref{fig:embedding0}
depicts the images at the same scale. The coordinate curves on the
hyperboloid are not those of embedding coordinates. These classic
embeddings are presented here to illuminate similarity; a formal
relationship is developed below.

\section{The Ambient Space}
\label{section:ambient}
\setcounter{equation}{0}
\setcounter{figure}{0}


Consider the ambient space $\C^3=\{(T,X,Y)\}$, and
write $T=T_1+iT_2$, etc. Endow~$\C^3$ with the flat metric
\[
E = \re(dT^2+dX^2+dY^2)
  = (dT_1^2 + dX_1^2 + dY_1^2) - (dT_2^2 + dX_2^2 + dY_2^2).
\]
Euclidean space is identified with the totally real slice
$\R^3\subset\C^3$, while Minkowski space is the 3-dimensional subspace
$i\R\times\R^2=\R^\Sig$ defined by the equations $T_1=X_2=Y_2=0$.  The
\emph{ambient Wick rotation} is the map
$\Wick:(T,X,Y)\mapsto(iT,X,Y)$, which carries $\R^3$~to $\R^\Sig$.

The ambient symmetry group that realizes the intrinsic symmetries of
our surface metrics is the additive complex group acting as
``(complex) rotation about the $Y$-axis'':
\begin{equation}
\label{eqn:action}
\Action_\theta(T,X,Y) =
(X\sin\theta + T\cos\theta, X\cos\theta - T\sin\theta, Y).
\end{equation}
A short calculation gives
\[
\Wick^{-1}\Action_{i\theta}\Wick(T,X,Y) =
(X\sinh\theta + T\cosh\theta, X\cosh\theta + T\sinh\theta, Y).
\]
This formula hints at the close relationship between rotations of
Euclidean space and boosts of Minkowski space.

Henceforth, the ambient space is essentially
$\C^2\times\R\subset\C^3$. In fact, the embeddings of primary interest
take values in $\C\times\R\times\R$; only~$T$ is ``really'' complex.
However, equation~\eqref{eqn:action} shows that continuous
interpolation of embeddings requires consideration of $\C^2=\R^{2,2}$.

\subsection*{Cylindrical Coordinates}

The \emph{complex cylindrical coordinates} mappings
$\cyl^\pm:\C^2\to\C^2$ are defined by
\begin{equation}
  \label{eqn:cylindrical}
  \begin{split}
  (T,X) &= \cyl^+(\rad,\Angle)= e^\rad(\sin\Angle, \cos\Angle) \\
  (T,X) &= \cyl^-(\rad,\Angle)=ie^\rad(\cos\Angle,-\sin\Angle)
  \end{split}
\end{equation}
Two mappings are required to cover a dense open set in~$\C^2$, see
below.  The ambient group action is coordinate translation:
\[
\Action_{\theta}\cyl^\pm(\rad,\Angle)=\cyl^\pm(\rad,\Angle+\theta).
\]


The metric quadratic form $Q=t_1^2+x_1^2-t_2^2-x_2^2$
partitions~$\C^2$ into sets of spacelike (for which~$Q$ is positive),
timelike (for which~$Q$ is negative), and lightlike vectors. Each of
these sets is a union of rays into the origin, so understanding these
sets amounts to understanding the way each intersects the unit sphere
$S^3=\{t_1^2+x_1^2+t_2^2+x_2^2=1\}$.  The lightlike cone intersects
the sphere in a real 2-torus whose complement consists of two linked
solid tori. The real cones over these solid tori are the sets of
spacelike and timelike vectors, and the mapping $(T,X)\mapsto(iX,-iT)$
(among others) involutively exchanges these sets. The image
of~$\cyl^+$ is the set of spacelike vectors, so the image of~$\cyl^-$
is the set of timelike vectors.

The pullback tensors $(\cyl^\pm)^*E$ express the metric~$E$ ``in
cylindrical coordinates'':
\begin{equation}
  \label{eqn:E-cyl}
  (\cyl^\pm)^*E =  \pm e^{2\rad}(d\Angle^2 + d\rad^2) + dY^2.
\end{equation}
It is straightforward to verify that the restrictions to the real
3-dimensional ambient slices of interest satisfy
\begin{align*}
  E^3    &=  e^{2\rad}(\Neg d\Angle^2 + d\rad^2) + dY^2, \\
  E^\Sig &=  e^{2\rad}(    -d\Angle^2 + d\rad^2) + dY^2.
\end{align*}
To check the second equation, it is necessary to use both cylindrical
coordinate regions.

An equivariant embedding of a surface is one for which (intrinsic)
$t$-translation is induced by $\Angle$-translation in cylindrical
coordinates.

\section{Embeddings}
\label{section:embeddings}
\setcounter{equation}{0}
\setcounter{figure}{0}

Fix a real constant $\grav>0$, let $\profile:(a,b)\to\R$ be a positive
profile satisfying $|\profile'|\leq2\grav$, and let $g$~be the metric
on $\C\times(a,b)$ as in~\eqref{eqn:metric_r}. The functions
\begin{equation}
  \label{eqn:embedding}
  \begin{split}
    T &= \frac{1}{\grav}\sqrt{\profile(r)}\sin(\grav t) \\
    X &= \frac{1}{\grav}\sqrt{\profile(r)}\cos(\grav t)\vphantom{\Bigg|}
  \end{split}
  \qquad
  Y = \int^r \sqrt{\frac{1}{\profile(\zeta)}\Bigl[1-
    \Bigl(\frac{1}{2\grav}\profile'(\zeta)\Bigr)^2\Bigr]}
    \,d\zeta
\end{equation}
define a mapping $f:\C\times(a,b)\to\C^3$. (The function~$Y$, which
depends only on~$r$, is well-defined up to an additive constant, which
alters the mapping by an ambient translation.)  Not incidentally,
$Y$~is real-valued.  To see that~$f$ is an isometric embedding of~$g$
into~$(\C^3, E)$, use cylindrical coordinates:
\[
d\Angle = \grav\,dt,\qquad
d\rad = \frac{\profile'(r)}{2\profile(r)}\,dr,\qquad
dY = \sqrt{\frac{1}{\profile(r)}\Bigl[1-
    \Bigl(\frac{1}{2\grav}\profile'(r)\Bigr)^2\Bigr]}\,dr
\]
so the pullback of~$E$ by~$f$ is
\begin{align*}
f^*E = e^{2\rad}(d\Angle^2+d\rad^2)+dY^2
  &= \frac{\profile(r)}{\grav^2}\Bigl(\grav^2\,dt^2
    +\frac{\profile'(r)^2}{4\profile(r)^2}\,dr^2\Bigr) + dY^2  \\
  &= \profile(r)\,dt^2+\frac{1}{\profile(r)}\,dr^2 = g.
\end{align*}
Observe that the embedding ``decouples'' in complex cylindrical
coordinates: $\Angle$~is a function of~$t$ alone, while $\rad$~and $Y$
are functions of~$r$ alone. Since the ambient and intrinsic group
actions are translations in $\Angle$~and $t$ respectively,
equivariance of~$f$ is obvious. 

\subsection*{Negative Profiles}

The embedding~\eqref{eqn:embedding} was introduced under the condition
that $\profile>0$. If $\profile<0$ on an interval, the wish that~$Y$
be real-valued imposes the condition $|\profile'|\geq2\grav$.  In this
situation, we are led to define
\begin{equation}
  \label{eqn:embedding-II}
  \begin{split}
    T &= \Neg\frac{1}{\grav}\sqrt{-\profile(r)}\cos(\grav t) \\
    X &=    -\frac{1}{\grav}\sqrt{-\profile(r)}\sin(\grav t)
                                              \vphantom{\Bigg|}
  \end{split}
  \qquad
  Y = \int^r \sqrt{\frac{1}{\profile(\zeta)}\Bigl[1-
    \Bigl(\frac{1}{2\grav}\profile'(\zeta)\Bigr)^2\Bigr]}
    \,d\zeta
\end{equation}
The formal expression of this mapping in terms of the cylindrical
coordinate mapping~$\cyl^-$ is identical to that
of~\eqref{eqn:embedding} with respect to~$\cyl^+$.  In particular,
equation~\eqref{eqn:embedding-II} defines an equivariant isometric
embedding of the metric
\[
g=\frac{1}{\profile(r)}\,dr^2+\profile(r)\,dt^2,\qquad \profile<0,
\]
into~$\C^3$.  Since $\profile<0$ but $t$~is imaginary, the intrinsic
coordinates have ``switched roles'': $r$~is timelike and $t$~is
spacelike.

\subsection*{The Schwarzschild-de Sitter Metric}

\begin{figure}[htb]
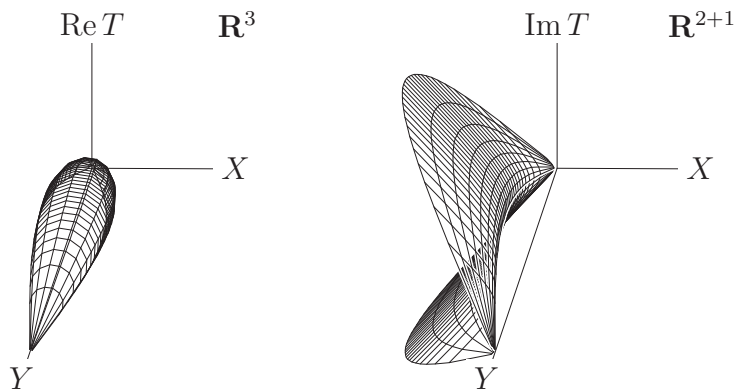

\begin{center}
\input{embedding2.eepic}\qquad
\input{embedding1.eepic}
\caption{A surface of revolution and the dual spacetime slice.}
\label{fig:embedding1}
\end{center}
\end{figure}

The metrics associated to the positive portion of the profile
$\profile(r)=1-\frac{2}{r}-\frac{r^2}{100}$ are depicted in
Figure~\ref{fig:embedding1}. The $X$-~and $Y$-axes are the real
axes of the complex coordinate directions.  The spacetime slice that
results is part of the Schwarzschild-de~Sitter metric.

Each surface intersects the $(X,Y)$-plane in the curve whose
parametrization may be read off equation~\eqref{eqn:embedding}; the
respective surfaces are swept out as this curve is acted on by a real
or imaginary coordinate translation in~$\Angle$.

\section{Fixed Points and Horizons}
\label{section:boundary}
\setcounter{equation}{0}
\setcounter{figure}{0}

In the previous section we omitted points at which $\profile=0$,
including fixed points of the group action and null lines in the
ambient space. In this section, we investigate issues related to
boundary conditions.

\subsection*{Riemannian Signature}

In the Riemannian situation, the profile is non-negative,
$\sqrt{\profile(r)}$~is the length of the Killing field generating the
group action, and a $t$-interval of length~$\frac{2\pi}{\grav}$
corresponds to one full turn of the image surface.  Consequently, the
action has a fixed point at each zero of~$\profile$. To understand the
significance of~$\grav$, we look more closely at the geometry of~$g$
near a fixed point.

Without loss of generality, we transform such that $\profile(0)=0$, and 
$\profile>0$ in some interval to the right of~0.  The Taylor expansion
is 
\[
\profile(r)=\profile'(0)r+o(r)\qquad
\text{near~$r=0$}.
\]
Let~$\theta$ be the cone angle at the vertex. Because the area element
of~$g$ is $dA=dr\,dt$, the portion of surface $0\leq r\leq\eps$ has
area~$2\pi\eps/\grav$. The boundary curve $\{r=\eps\}$ has
circumference $2\pi\sqrt{\profile(\eps)}/\grav$.  By elementary
geometry, the cone angle is 
\[
\theta
=\lim_{\eps\to0}\frac{\mathrm{circumference}^2}{2\cdot\mathrm{area}}
=\frac{\pi}{\grav}\profile'(0).
\]
Since the image of the embedding is smooth at a fixed point iff
$\theta=2\pi$, smoothness is equivalent to $\profile'(0)=2\grav$. A
similar argument holds if~$\profile$ is positive to the left of~0: The
embedding is smooth iff $\profile'(0)=-2\grav$. Generally,
$\profile'(0)$~is a measure of angular defect (a.k.a.\ point
curvature) at a fixed point.

In~\cite{H}, the value $\grav=1$ was used tacitly. If~$\grav\neq1$, it
is appropriate to view the embedding~\eqref{eqn:embedding} as taking
values not in~$\R^3$, but in a space that is flat away from the
$Y$-axis and has cone angle~$2\pi/\grav$ along the $Y$-axis.  If
$|\profile'|=2$ at a zero of the profile, the embedding meets the
$Y$-axis perpendicularly; the intrinsic and ambient metric
singularities have the same total angle.

\subsection*{Lorentz Signature}

\begin{figure}[hbt]
\begin{center}
\input{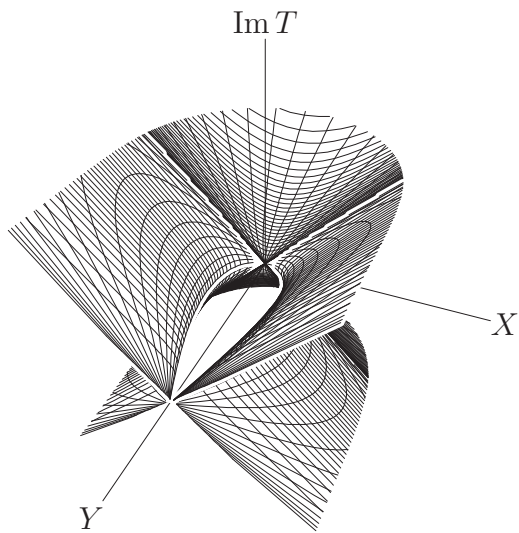}
\smallskip
\caption{Extension of the Schwarzschild-de~Sitter embedding}
\label{fig:embedding3}
\end{center}
\end{figure}

Provided the radicand in the definition of~$Y$ is non-negative, the
embedding of a spacetime slice extends to intervals on
which~$\profile$ achieves negative values. Geometrically, a zero of
the profile must map to a null line in~$\R^\Sig$, and the image of the
embedding can cross from the spacelike portion of~$\R^\Sig$ to the
timelike portion, see Figure~\ref{fig:embedding3}.

A \emph{horizon} of an embedding is a curve in~$\R^\Sig$ along which
the image is null. Intrinsically, a horizon comes from a zero
of~$\profile$, where the spacetime slice degenerates.  A glance
at~\eqref{eqn:embedding} shows that the embedding is smooth at a
horizon only if $|\profile'|=2\grav$ and $\profile''<0$.
There are two horizons in Figure~\ref{fig:embedding3}; one in the
foreground (a cosmological horizon), and one containing the origin
(the black hole horizon).

\subsection*{Embeddability Criteria}

Engman~\cite{Engman} studied momentum profiles and their associated
surfaces of revolution, and characterized embeddability in terms of
bounds on the total curvature of a disk centered at a fixed point.
His bounds naturally have a spacetime analogue, though a satisfactory
topological interpretation is lacking due to non-compactness.

The portion of metric associated to a profile~$\profile$ embeds iff
the coordinate function~$Y$ in~\eqref{eqn:embedding} is real-valued.
Non-negativity of the radicand is an obvious necessary condition, and
is sufficient if the profile is (say) real-analytic at each zero. If a
metric associated to~$\profile$ embeds, then at each point of the
momentum interval one of the following conditions holds:

\bi

\item $\profile>0$ and $|\profile'|\leq2\grav$,

\item $\profile<0$ and $|\profile'|\geq2\grav$,

\item $\profile=0$, $|\profile'|=2\grav$.

\ei

In the Riemannian situation, the profile must be non-negative, and the
metric embeds iff $|\profile'|\leq2\grav$. To obtain bounds on the
total curvature of a disk, assume without loss of generality that
$\grav=1$ (so that the ambient space is smooth), $\profile(0)=0$, and
$\profile'(0)=2$. For $r_1>0$, the total curvature of the disk
$D=\{0\leq r\leq r_1\}$ is
\[
\int_D \curv\,dA=-\pi\int_0^{r_1}\profile''(r)\,dr
=\pi\bigl(2-\profile'(r_1)\bigr).
\]
We read off at once that the metric embeds iff the total curvature of
an arbitrary disk (centered at the fixed point) is between $0$~and
$4\pi$. A similar analysis holds for profiles that are positive to the
left of a zero. 

The upper bound is topologically significant: By the Gauss-Bonnet
theorem, a surface of revolution generated by a profile that vanishes
twice has total curvature~$4\pi$, even if the surface is not smooth at
a fixed point. There are two ways the upper bound can be achieved by a
profile satisfying $\profile(0)=0$ and $\profile'(0)=2$: The profile
can become too steep (downward), or vanish at some $r_1>0$. In either
case, $\profile$~no longer defines an embedding.

The spacetime situation is analogous to an extent: There is an
embeddability criterion in terms of total curvature, which has the
expected sign change from the Riemannian case. Because the symmetry
group is non-compact, we must speak of curvature \emph{per unit time}.
Further, it is meaningful to speak of embeddings defined by negative
profiles, so there are two cases to consider. As above, we
assume~$\profile$ is a profile satisfying $\profile(0)=0$~and
$\profile'(0)=\pm 2\grav$, and that~$\profile$ is non-vanishing between
$0$~and $r_1$ for some $r_1>0$.

If $\profile>0$, embeddability is equivalent to the pointwise
inequality $|\profile'|\leq2\grav$. Integrating the curvature
form~$\curv\,dA$ over a $t$~interval of length~1 and the $r$~interval
$[0,r_1]$, we find that the total curvature per unit time is
non-positive, but no less than~$-4\pi$, in every neighborhood of the
horizon $r=0$.

If $\profile<0$, embeddability is equivalent to the bounds
$|\profile'|\geq2\grav$. By an entirely analogous argument, the total
curvature over~$[0,r_1]$ is either $\geq2\pi$ or is non-positive.
However, total curvature is a continuous function of~$r_1$, and since
the total curvature vanishes in the limit as $r_1\to0$, the first
bound cannot hold for arbitrary~$r_1$. 

We conclude that embeddability of a piece of spacetime metric
associated to a negative profile implies that the metric has
non-positive total curvature in every neighborhood of the horizon,
regardless of the sign of the profile. In an interval where the
profile is positive, the total curvature is no smaller than~$-4\pi$,
while if the profile is negative, there is no \emph{a~priori} lower
bound on the total curvature.

Non-positivity of the total curvature implies the observation made
in~\cite{GMG}: if $\profile'(0)=2\grav\neq0$ then the metric embeds in
a neighborhood of the horizon if $\profile''<0$.

\end{document}